\documentclass{aastex}
\usepackage{emulateapj5}
\usepackage{onecolfloatx}

\begin{document}

\twocolumn[%

\journalinfo{astro-ph/0010489}
\title{On the Mass and Inclination of the PSR~J2019+2425 Binary System}

\author{David J. Nice and Eric M. Splaver}
\affil{Physics Department, Princeton University \\ Box 708, Princeton, NJ  08544}
\email{dnice@princeton,edu, esplaver@princeton.edu}

\and

\author{Ingrid H. Stairs$^1$}
\affil{Jodrell Bank Observatory, University of Manchester \\ Jodrell Bank,
Macclesfield, Cheshire, SK11 9DL, United Kingdom}
\email{istairs@gb.nrao.edu}

\medskip

\submitted{Accepted by the Astrophysiccal Journal, 24 October 2000}

\begin{abstract}
We report on nine years of timing observations of PSR~J2019+2425,
a millisecond pulsar in a wide 76.5~day orbit with a white dwarf.  We measure
a significant change over time of the projected semi-major axis of the orbit,
${\dot{x}/x}=1.3\pm0.2\times 10^{-15}$\,s$^{-1}$, where $x\equiv(a_1\sin i)/c$.
We attribute this
to the proper motion of the binary.  This constrains the inclination angle
to $i<72^\circ$, with a median likelihood value of $63^\circ$.  
A similar limit on inclination angle arises
from the lack of a detectable Shapiro delay signal.  These limits
on inclination angle, combined with a model of the evolution of
the system, imply that the neutron star mass is at most 1.51\,$M_\odot$;
the median likelihood value is 1.33\,$M_\odot$.
In addition to these timing results, we present
a polarization profile of this source.  Fits of the linear polarization
position angle to the rotating vector model indicate the magnetic axis
is close to alignment with the rotation axis, $\alpha<30^\circ$.

\end{abstract}


]

\section{Introduction}

\begin{figure}[b]
{\footnotesize
$^1$Present address: NRAO, Box 2, Green Bank, WV 24944}
\end{figure}
\addtocounter{footnote}{1}

Neutron star masses measured in radio pulsar binary systems are
consistent with a remarkably small range, \mbox{$m=1.35\pm 0.04$\,M$_\odot$}
 (Thorsett \& Chakrabarty 1999)\nocite{tc99}.  The best such
measurements, those with uncertainty less than 5\%, are of pulsars in tightly
bound, highly eccentric neutron star--neutron star binaries.  In such systems,
relativistic phenomena detected in radio pulse timing experiments allow the direct,
high precision measurement of the masses of the component stars.  By contrast,
measurement of masses of pulsars in neutron star--white dwarf systems tend to
have lower precision, or to be statistical in nature.  Measurement of masses
in these systems are of interest because their evolution is substantially
different from that of neutron star--neutron star binaries.  In particular,
neutron star--white dwarf systems go through an extended period of mass
transfer, during which the secondary loses several tenths of a solar mass of
matter, at least some of which is accreted onto the neutron star, making the
system visible as a low mass X-ray binary.  Thus, one might expect the neutron
stars in these systems to be more massive than those in neutron star--neutron
star binaries.

In this paper, we describe pulse timing observations of PSR~J2019+2425, a
millisecond pulsar in a 76.5~day orbit with a white dwarf.  Our observations
constrain the ({\it a priori} unknown) inclination angle of the orbit.
By combining this constraint with a theory of
orbital evolution, we determine an upper limit of the mass of the pulsar.

We have previously reported on this pulsar in Nice \& Taylor
(1995)\nocite{nt95}.  The present work represents a tripling
of the time span of the observations.  In \S\ref{sec:obs} we
describe the data acquisition.  In \S\ref{sec:timing} we present
an analysis of the pulse arrival times.
Implications for the neutron star mass are given in \S\ref{sec:mass}.
Polarimetry of the pulsar is discussed in \S\ref{sec:polar}.

\bigskip

\section{Observations}\label{sec:obs}

\subsection{Data collection}

\begin{figure*}[t]
\epsscale{2.1}
\plotone{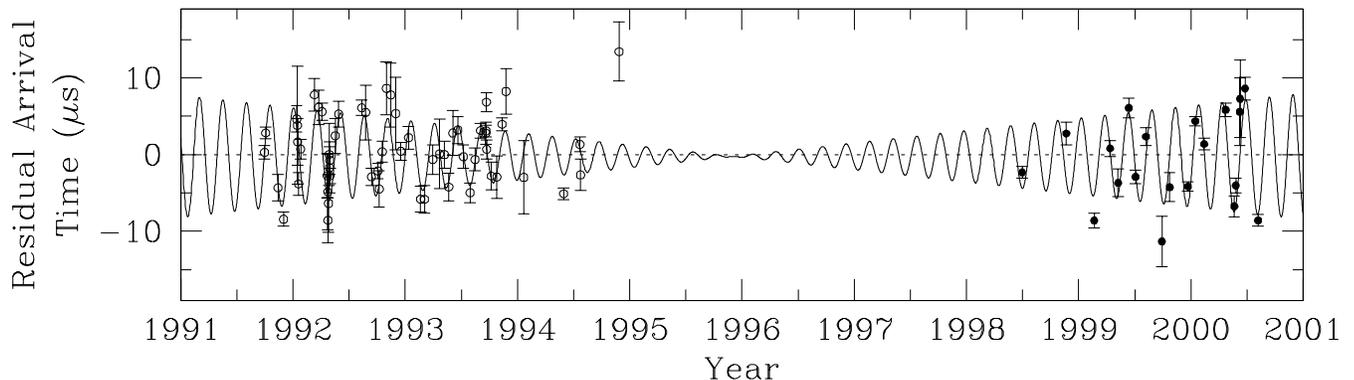}
\epsscale{1}
\caption{Residual pulse arrival times after removing the 
timing model in table~\ref{tab:param} but with constant projected semi-major
axis ($\dot{x} \equiv 0$).  Open and closed circles indicate Mark~III and
Mark~IV data, respectively.  The solid line shows the residual arrival times expected from
the changing apparent orbital size, $\dot{x}=5\times 10^{-14}$; this curve is
sinusoidal with the orbital period with amplitude increasing linearly
towards the ends of the data set.
Arrival time uncertainties shown in the figure 
do not include corrections for systematics.
\label{fig:resid}}
\end{figure*}

We measured times of arrival (TOAs) of pulses from PSR~J2019+2425 on 78
separate days between 1~October~1991 and 7~August~2000, using the 305\,m radio
telescope at Arecibo.  We made all observations at 430\,MHz, primarily with
the telescope's line feed, except for a few of the most recent observations,
for which we used the Gregorian 430\,MHz receiver.

We employed two distinct data acquisition systems.  Between October 1991 and
November 1994, we used the Princeton Mark III data acquisition system
(Stinebring {\it et al.} 1992)\nocite{skn+92}.  An analog filter bank divided
an 8~MHz passband into thirty-two 0.25~MHz spectral channels in each sense of
circular polarization.  These signals were detected with 100~$\mu$s time
constants, and opposite polarizations were summed.  The resulting intensity
levels were sampled and summed synchronously with the pulsar period.  Pulse
profiles were accumulated for integration times of 2--3 minutes and stored for
later analysis.  Off-line, profiles from different spectral channels were
shifted in time to compensate for interstellar dispersion and summed to
produce a single de-dispersed profile for each integration.  A combination of
differential dispersion smearing within the spectral channels and analog
filter bank time constants limited the time resolution of these profiles to
$450\mu$s.

No observations were made between November 1994 and June 1998 due to
the Arecibo telescope upgrade project.

Between July 1998 and August 2000, observations were made as part of the
Arecibo coordinated timing project.  We used the Princeton Mark IV system for
data collection \cite{sst+00}.  In each sense of circular polarization, a
5~MHz passband was mixed to baseband using local oscillators in phase
quadrature.  The four resulting signals were low-pass filtered at 2.35\,MHz,
sampled at 5\,MHz, quantized to 4 bits, and stored on disk or tape.  Upon
playback, these voltages were coherently dedispersed using software
techniques.  Self- and cross-products of the complex voltages were calculated,
giving power measurements in four Stokes parameters.  These were summed
synchronously with the pulse period with integration times of 190 seconds,
yielding pulse profiles with full polarimetry.  The profiles were calibrated
by measurement of a pulsed noise calibration signal injected at the telescope
receiver at the start or end of each observation; the noise calibration
signals themselves were calibrated against continuum radio sources.

\subsection{Time of arrival measurements}

We derived TOAs by fitting a standard template to the total intensity
de-dispersed profiles.  Separate standard templates were used for
the Mark~III and Mark~IV data sets.  The locations of the pulse peak 
measured in the fit
were added to the scan start times to produce TOAs.  For pre-upgrade
observations, the start times were referenced to an observatory clock, which
we corrected to UTC(NIST) and, ultimately, UTC. For post-upgrade observations,
start times were referenced to GPS time, which tracks UTC within tens of
nanoseconds, sufficient precision for our purposes.

We fit for an arbitrary offset between the Mark~III and Mark~IV TOAs.  The
measured offset, $43\pm3\,\mu$s, results from a combination of differences in
signal propagation time through the hardware and differences in the standard
templates used for measuring TOAs.

The set of TOAs from each day of observations were combined into a single
effective TOA for that day.  Days with very high uncertainties or few
measurements were eliminated.  The final data set consisted of 78 TOAs,
of which 58 were taken with Mark~III (1991--1994) and 20 were taken 
with Mark~IV (1998--2000).

\subsection{TOA uncertainties}

An uncertainty was calculated for each daily TOA based on the scatter of the
arrival times within that day.  The root-mean-square of the uncertainties thus
calculated was 1.4\,$\mu$s for both the Mark III and Mark IV data.  (While the
time resolution of the Mark IV data was much higher than that of the Mark III
data, the Mark~III observations had a larger bandwidth and, usually, a
substantially longer on-source time, resulting in similar overall precision.)
In the timing analysis described below, we found the $\chi^2$ values to be
systematically high, $\chi^2/\nu=3.3$, where $\nu$ is the number of degrees of
freedom of the fit.  There was no evidence of long-term or orbit-dependent
trends in the pulse arrival times, suggesting the high $\chi^2$ values are
day-to-day jitter in our measurements, or systematic underestimates of our
uncertainties, of unknown origin.  We compensated for this by adding
$2.4\,\mu$s in quadrature to the statistical uncertainty calculated on each
day, which yielded fits with $\chi^2/\nu\approx 1$.  By varying this systematic
term independently for the Mark III and Mark IV data while analyzing separate
$\chi^2$ statistics for these two subsets of the data, we found that similar
quadrature terms were needed for both subsets.

\begin{table*}[t]
\caption{Parameters of PSR~J2019+2425 System\label{tab:param}\tablenotemark{a}}
\centerline{
\begin{tabular}{ll}
\hline
\hline
Right ascension, $\alpha$ (J2000)\dotfill & $20^{\rm h}19^{\rm m}31\fs94900(3)$ \\
Declination, $\delta$ (J2000)\dotfill     & $+24\arcdeg25\arcmin15\farcs3038(5)$ \\
Proper motion in R. A., $\mu_{\alpha *}=(d\alpha/dt)\cos\delta$ (mas/yr)\dots   
                                          & $-9.41$(12)                \\
Proper motion in Dec., $\mu_\delta=d\delta/dt$ (mas/yr)\dotfill    
                                          & $-20.60$(15)               \\
Period (ms)\dotfill                       & 3.93452408033124(11)     \\
Period derivative ($10^{-21}$)\dotfill    & 7.0237(12)               \\
Epoch (MJD)\dotfill                       & 50000.0000               \\
Orbital period, $P_b$ (days) \dotfill                & 76.51163479(2)        \\
Change rate of orbital period, $\dot{P_b}$ \dotfill  & $-3(6)\times 10^{-11}$  \\
Projected semi-major axis, $x$ (lt-s)\dotfill        & 38.7676297(8)         \\
Change rate of projected semi-major axis, $\dot{x}$ \dotfill      & $5.1(8)\times 10^{-14}$ \\
Eccentricity\dotfill                      & 0.00011109(4) \\
Longitude of periastron, $\omega$\dotfill & $159\fdg03(2)$ \\
Time of periastron passage\tablenotemark{b}\ (MJD)\dotfill  & $50054.6439021\pm0.004$ \\
Dispersion measure\tablenotemark{c}\ (pc\,cm$^{-3}$)\dotfill  &  17.203 \\
Mass function, $f_1$ ($M_\odot$) \dotfill & 0.0106865005(6) \\
\hline
\multicolumn{2}{l}{\footnotesize $^a$Values in parentheses are ``$2\sigma$'' uncertainties 
(95\% confidence) in the last digit quoted.} \\
\multicolumn{2}{l}{\footnotesize $^b$Highly covariant with longitude of
periastron;  value corresponds to $\omega=159\fdg0300000.$}\\
\multicolumn{2}{l}{\footnotesize $^c$Held fixed at the value found by Nice, Fruchter, \& Taylor
(1993)\nocite{ntf93}.} \\
\end{tabular}
}
\end{table*}

\section{Timing analysis}\label{sec:timing}
\subsection{Timing model and $\dot{x}$}\label{sec:xdot}

We used the {\sc Tempo}\footnote{http://pulsar.princeton.edu/tempo} 
software
package to fit a pulse timing model to the observed TOAs.  A
standard timing model incorporating spin-down (period and period derivative),
astrometric parameters (position and proper motion), and five Keplerian
parameters of the pulsar orbit (orbital period, semi-major axis projected into
the line of sight, eccentricity, angle of periastron, and time of periastron
passage) is not adequate to fully describe the observed TOAs.  
It is necessary to also allow a secular change in the projected
semi-major axis,
 $\dot{x}=dx/dt$, where $x\equiv(a_1\sin i)/c$, and
$a_1$ is the semi-major axis of the pulsar orbit, $i$ is the inclination of
the angular momentum vector of
the orbit relative to the earth--pulsar line-of-sight, $c$ is the speed of
light, and the dot indicates a
time derivative.
The need for a nonzero $\dot{x}$ is illustrated in
figure~\ref{fig:resid}.
The measured value of $\dot{x}$ is
\begin{equation}
\dot{x}=(5.1\pm 0.8)\times 10^{-14}.
\end{equation}
The full set of parameter values from the best timing model fit is listed 
in table~\ref{tab:param}.  Uncertainties were calculated by a bootstrap
procedure.  Values given in the table are twice the formal uncertainties.
Since all observations were made at a single radio frequency, the dispersion
measure was held fixed in the timing analysis.

\subsection{Interpretation of $\dot{x}$}\label{sec:xdottheory}

The nonzero $\dot{x}=(1/c)\,d(a_1\sin i)/dt$ could, in principle, result from
a change in orbital size, $a_1$, or inclination angle, $i$,
or a combination.  First we will consider (and reject) the possibility
of a change in $a_1$.  Then we will analyze the implications of a 
change in inclination angle.

Peters (1964)\nocite{pet64} calculates the change
in orbital size of a system of two point masses under general relativity,
\begin{eqnarray}
\lefteqn{\dot{a_1}=-\frac{64}{5}T_{\odot}^3\frac{m_1m_2^5}{(m_1\!+\!m_2)^3}\frac{1}{a_1^3}
\frac{1}{\left(1-e^2\right)^{7/2}}\times}\hspace*{80pt} \nonumber\\
& & \times \left(1+\frac{73}{24}e^2+\frac{37}{96}e^4\right),
\end{eqnarray}
where $T_{\odot}\!=\!GM_\odot/c^3\!=\!4.925\!\times\!10^{-6}\,$s, $m_1$ and
$m_2$ are the pulsar and white dwarf masses expressed in solar masses, $e$ is
the eccentricity, and $a_1$ is expressed in light-seconds.  While $m_1$,
$m_2$, and $a_1$ are not unambiguously known, we can estimate the magnitude of
$\dot{a_1}$ by using \mbox{$m_1\sim1.4\,$M$_\odot$}, \mbox{$m_2\sim0.35\,$M$_\odot$}, and
\mbox{$a_1\sim 40$\,lt-sec}, from which \mbox{$\dot{a_1}\sim3\times 10^{-23}$}.  
This is many orders
of magnitude below the observed \mbox{$\dot{x}=5\times10^{-14}$}.

More generally, for typical astrophysical processes within a binary system,
$|\dot{a_1}/a_1|$ will be the same order of magnitude as $|\dot{P_b}/P_b|$,
where $P_b$ is the orbital period.  Interpreting the observed value of
$\dot{P_b}$ (table~\ref{tab:param}) as an upper limit, $|\dot{P_b}|<9\times
10^{-11}$, we have $|\dot{P_b}/P_b|<1.4\times 10^{-17}$.  This is two orders
of magnitude smaller than the observed $|\dot{x}/x|=1.3\times 10^{-15}$, so we
conclude that the observed nonzero $\dot{x}$ is not caused by orbital
evolution.

\begin{figure}[b]
\epsscale{0.8}
\plotone{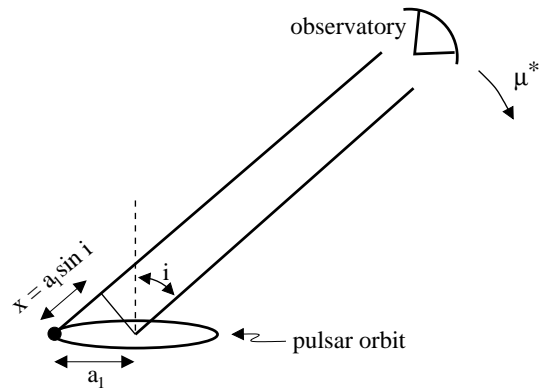}
\epsscale{1}
\caption{Geometry relevant to the observation of $\dot{x}$.  
See section \S~\ref{sec:xdottheory}.\label{fig:pm}}
\end{figure}

The nonzero $\dot{x}$ must arise because of a change in the observed
inclination of the orbit.  Kopeikin (1996)\nocite{kop96} discusses how
apparent changes in orbital parameters arise due to the relative motion of the
binary and the observer.  The situation is sketched in figure~\ref{fig:pm}.  A
component of the proper motion lies in the plane formed by the line-of-sight
to the binary and the angular momentum vector of the binary.  We write this as
$\mu^*=\mu\sin\theta$, where $\mu$ is the total proper motion and
$\theta=\theta_\mu-\Omega$ is the difference between the position angle of
proper motion, $\theta_\mu$, and the position angle of the ascending node,
$\Omega$.  It is clear that this component of proper motion equals the change
in inclination angle, $\mu^*=di/dt$ (figure~\ref{fig:pm}).  From the
definition of $x$,
\begin{equation}\label{eqn:coti}
\frac{\dot{x}}{x}=\mu\,\cot i\,\sin\theta.
\end{equation}

The orientation of the binary, $\theta$, is not known.  However, requiring
$|\sin\theta|<1$ in equation~\ref{eqn:coti} gives a firm upper limit on $i$.
Using $\mu=22.62\pm0.15$\,mas/yr, $x=38.7676$\,lt-s, and $\dot{x}=4.3\times
10^{-14}$ (the lower limit of allowed values from table~\ref{tab:param})
in equation~\ref{eqn:coti} yields
\begin{equation}
i<\tan^{-1}\left(\mu\frac{x}{\dot{x}}\right)=72^\circ.
\end{equation}

The distribution of inclination angles within this constraint can be studied
using Monte Carlo analysis.  We make the {\it a priori} assumption that all
orientations of the binary system in space are equally likely.  Under this
assumption, $\theta$ is a uniformly distributed random variable and $i$ is a
random variable distributed with uniform probability in $\cos i$.  We select
values of $\theta$ and $i$ from these distributions and retain only those
combinations which satisfy equation~\ref{eqn:coti} within the measurement
uncertainties.  In the resulting data set, the distribution of $i$ is somewhat
peaked towards the highest allowed values.  The median likelihood 
inclination angle
within this distribution is $63^\circ$.  We discuss the implications of this
in \S\ref{sec:mass}.

\subsection{Shapiro delay}\label{sec:shapiro}

According to general relativity, the pulse signal is delayed as it propagates
through the gravitational potential well of the secondary.  This ``Shapiro
delay'' for a pulsar in a circular orbit is
\begin{equation}
\Delta t = -2T_\odot m_2\ln[1-\sin i \sin \phi_b],
\end{equation}
where $\phi_b$ is the orbital phase
measured from the ascending node.  For small inclination angles, the variation
of $\Delta t$ over the orbit is nearly sinusoidal, so it is indistinguishable
from a slight increase in the ({\it a priori} unknown) orbital size, $a_1$.
For edge-on orbits ($i\approx 90^\circ$), the variation in $\Delta t$ becomes
strongly peaked at $\phi_b\approx 90^\circ$, when the pulsar is behind the
secondary; this breaks the covariance with the
Keplerian orbital model and allows measurement of the Shapiro delay (and hence
measurement of $m_2$ and $i$).

\begin{figure}[b]
\epsscale{1.}
\plotone{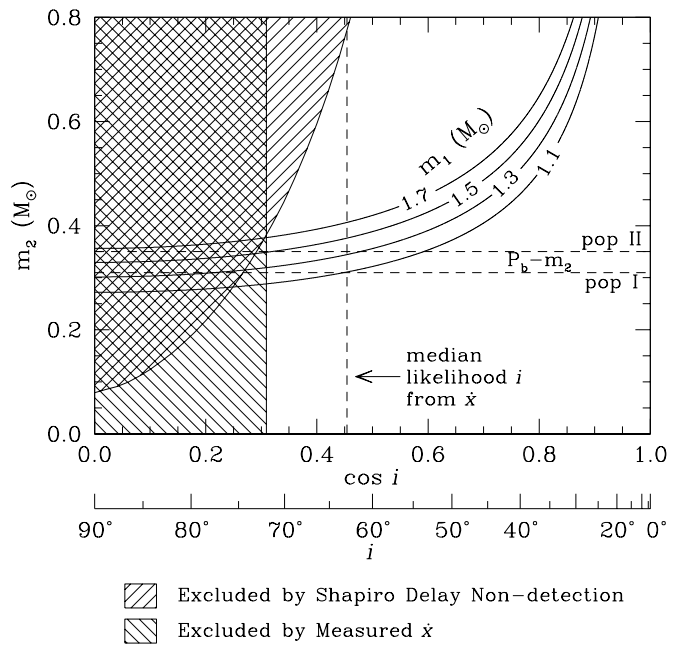}
\epsscale{1}
\caption{Constraints on the inclination angle, $i$, and white dwarf mass, $m_2$,
from measured $\dot{x}$ and from limits on the Shapiro delay.  Any point outside
the shaded regions is allowed by the timing data.  The ``median likelihood''
inclination angle, indicated by a dashed vertical line, is based on an {\it a
priori} uniform distribution in the direction of the orbital angular momentum
vector.  The horizontal dashed lines indicate the companion mass values
expected from the $P_b-m_2$ relation for population I and II secondaries.  The
solid curves indicate contours of constant neutron star mass, $m_1$, calculated
using the observed mass function for a given value of $i$ and $m_2$.
\label{fig:cosim2}}
\end{figure}

We did not detect the Shapiro delay in the PSR~J2019+2425 TOAs.  We can use
this non-detection to place limits on allowed inclination angles.  We analyzed
the TOAs of PSR~J2019+2425 using a grid of timing models, each of which
incorporated the Shapiro delay signal appropriate for some inclination angle
$0^\circ<i<90^\circ$ and secondary mass $0<m_2<0.8$\,M$_\odot$.  The fits to
most such models yielded $\chi^2$ values similar to a model with no Shapiro
delay.  However, models with high inclination angles and/or high secondary masses
fared poorly---meaning that the Shapiro delay would have been detected had
these been the true values of $i$ and $m_2$.  Figure~\ref{fig:cosim2} shows
the parameter space excluded by models with $\Delta\chi^2>4$ (i.e., $2\sigma$)
relative to the base model.  For modest secondary masses, the inclination
angle limit is $i\lesssim 73^\circ$, very similar to the limit from the
$\dot{x}$ measurement.

\section{Mass}\label{sec:mass}

The observational limits on the inclination angle, combined with binary
evolution theory, constrain the mass of the pulsar.  Like all wide neutron
star--white dwarf binaries, the PSR~J2019+2425 system underwent an extended
period of mass transfer, during which the pulsar was spun up to its short
period.  A straightforward model of this mass transfer gives a unique relation
between the orbital period, $P_b$, and secondary mass, $m_2$ at the end of
this phase of evolution (see Rappaport {\it et al.}  1995\nocite{rpj+95} and
references therein.).  Key elements of this model include (i) stable mass
transfer, with the secondary filling its Roche lobe in the giant phase; (ii)
well-defined dependence of the radius of the giant secondary on its core mass;
and (iii) dissipation of almost all of the outer envelope of the secondary,
so that the present day secondary mass, $m_2$, is only slightly more than the
secondary core mass.

In a numerical analysis of systems which evolve into neutron star--white dwarf
binaries, Tauris and Savonije (1999)\nocite{ts99a} found the relation between
orbital period (in days) and secondary mass (in solar masses) to be
\begin{equation}
m_2=\left(\frac{P_b}{b}\right)^{1/a}+c,
\end{equation}
where $a$, $b$, and $c$ depend on the chemical composition
of the secondary,
\begin{equation}
(a,b,c)=\left\{ 
\begin{array}{llll}
4.50 & 1.2\times 10^5 & 0.120 & \mbox{Pop. I} \\
4.75 & 1.1\times 10^5 & 0.115 & \mbox{Pop. I+II} \\
5.00 & 1.0\times 10^5 & 0.110 & \mbox{Pop. II}.\\
\end{array}
\right.
\end{equation}
For the orbital period of PSR~J2019+2425, $P_b=76.512$\,d, this yields masses
ranging from $m_2=0.31$\,M$_\odot$ for a population I donor to
$m_2=0.35$\,M$_\odot$ for a population II donor.  These values are indicated
in figure~\ref{fig:cosim2}.  In a similar analysis, Rappaport {\it et al.}
(1995)\nocite{rpj+95} found $0.26<m_2<0.35$.

The pulsar is likely very old. Its characteristic age, calculated after
correcting the observed period derivative for bias due to the translational
Doppler effect,
is 27\,Gyr (see the discussion in Nice \& Taylor 1995\nocite{nt95}).
Optical observations of the white dwarf secondary (Lundgren, Foster, \& Camilo
1996)\nocite{lfc96} give a cooling age of 7.6 to 13.9\,Gyr, assuming it is a
helium core \cite{hp98b}.  Thus the population~II value of
$m_2=0.35$\,M$_\odot$ is likely to be appropriate for this system.

The observed mass function of the pulsar is
\begin{equation}
f_1\equiv\frac{(m_2 \sin i)^3}{(m_1+m_2)^2}
        = \frac{4\pi^2c^3}{G}\frac{x^3}{P_b^2}
        = 0.010685\,{\rm M}_\odot.
\end{equation}
For $m_2=0.35$\,M$_\odot$ and $i<72^\circ$, this constrains the mass of 
the pulsar to be
\begin{equation}
m_1<1.51\,{\rm M}_\odot
\end{equation}
(see figure~\ref{fig:cosim2}).  This limit is conservative, in that
it holds not just for a population~II companion but for any secondary mass 
$m_2<0.35$\,M$_\odot$, i.e.,
the full range of masses considered by Rappaport {\it et al} (1995) and
Tauris \& Savonije (1999).  
For the particular case of $m_2=0.35$\,M$_\odot$, at the median likelihood value 
of $i=63^\circ$, the pulsar mass is
$m_1=1.33$\,M$_\odot$.

\section{Polarimetry}\label{sec:polar}

\begin{figure}[t]
\epsscale{1.}
\plotone{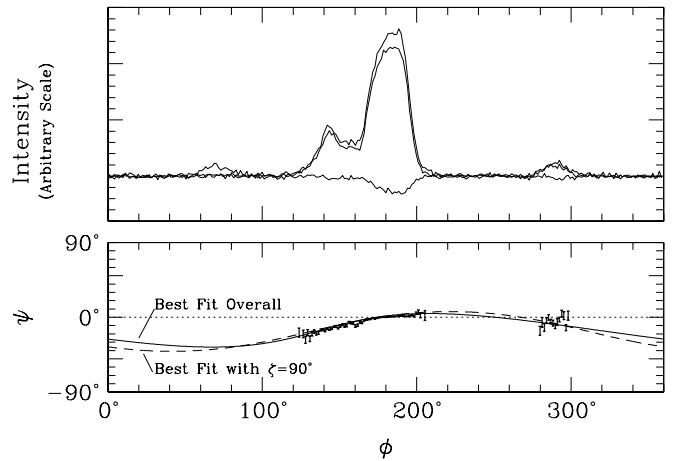}
\epsscale{1}
\caption{Pulse profile of PSR~J2019+2425 at 430~MHz.  The upper plot
shows total intensity (upper trace), linearly polarized intensity (middle trace),
and circularly polarized intensity (lower trace) as a function of pulse phase, $\phi$.
The lower plot shows the position angle of linear polarization, $\psi$.
(Position angle is defined as increasing counterclockwise on the sky, 
with an arbitrary
zero point.)
The solid curve in the lower plot shows the overall best fit of the
rotating vector model.  The dashed curve shows the best RVM fit
when restricted to $\zeta=90^\circ$.  The RVM fits were performed
on a 1024-bin profile, but a 256-bin profile is shown here for clarity.
\label{fig:prof}}
\end{figure}

\begin{figure}[b]
\plotone{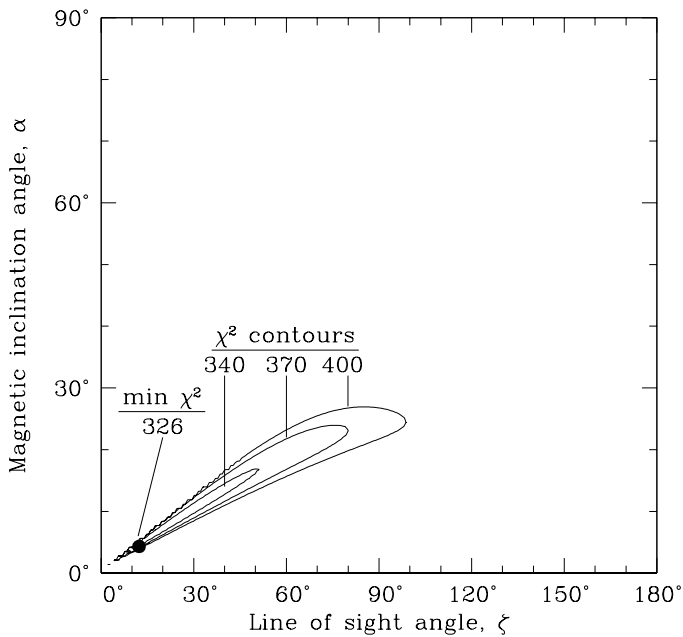}
\caption{Results of fits of the rotating vector model to the 
linearly polarized pulse profile.  Best-fitting values of
the angle between the rotation and magnetic axes, $\alpha$, 
and the angle between the rotation axis and the line of
sight, $\zeta$, are shown.  Representative $\chi^2$ values
are given from fits with 268 degrees of freedom.
\label{fig:az}}
\end{figure}

Since the angular momentum of the spun-up pulsar is almost entirely due
to mass transferred from the orbital companion, the angular momentum
vector of the pulsar must be aligned with that of the binary system.  Measurement
of the geometry of the pulsar could, therefore, give the inclination of the
orbit.  The standard picture of pulsar radio emission along dipole magnetic field
lines, the ``rotating vector model'' (RVM), yields a well known expression for
the position angle of linearly polarized radiation, $\psi$, as a function of
pulse phase, $\phi$:
\begin{equation}
\psi-\psi_0 = \frac{\sin\alpha \sin(\phi-\phi_0)}
        {\sin\zeta\cos\alpha-\cos\zeta\sin\alpha\cos(\phi-\phi_0)},
\end{equation}
where $\alpha$ is the angle between the rotation and magnetic axes of the
pulsar, $\zeta$ is the angle between the rotation axis and the line of sight,
and $\phi_0$ and $\psi_0$ are the pulse phase and position angle at the
point with the steepest change in position angle.
We expect $\zeta=i$ or $\zeta=180^\circ-i$ for pulsar rotation aligned
with the orbital angular momentum vector.

As discussed in \S\ref{sec:obs}, our Mark~IV data includes full polarization
information.  Figure~\ref{fig:prof} shows a polarization profile of
PSR~J2019+2425 created by summing all strong Mark~IV integrations.  The pulsar has
emission over a substantial fraction of the period and is highly linearly
polarized.  (An exception is that the weak component at $\phi=60^\circ$ is
unpolarized.)
Unfortunately, there is only modest variation in the position angle.
This is indicative of alignment between the rotation and magnetic axes
of the pulsar (small $\alpha$), and in practice makes it difficult to
discern $\zeta$.  We performed a grid search in $\zeta$ and $\alpha$,
finding the best RVM fit for each combination of these angles.  The
best overall fit had $\alpha=4^\circ$ and $\zeta=12^\circ$ with a
$\chi^2=326$ for 268 degrees of freedom; this fit is plotted in
figure~\ref{fig:prof}. This value of $\zeta$, $12^\circ$, is
surprisingly small, and would imply a very small neutron star mass
(figure~\ref{fig:cosim2}).  Though the formal significance of this fit
is reasonably good, RVM fits with values of $\zeta$ ranging from
$0^\circ$ to $90^\circ$ can be found with only a modest increase in
$\chi^2$, as shown in figure~\ref{fig:az}.  The best RVM fit with
$\zeta=90^\circ$ is also shown in figure~\ref{fig:prof};
qualitatively, it does not appear very different from the overall
best-fit RVM model.  It should also be noted that millisecond pulsars
often show deviations from the rotating vector model (e.g., Xilouris
{\it et al.}, 1998\nocite{xkj+98}; Stairs, Thorsett \& Camilo,
1999\nocite{stc99}).  Because of this, we conclude that the data
cannot be used to place tight limits on $\zeta$, although within the
context of the RVM we can say with confidence that $\alpha\lesssim
30^\circ$.

\section{Conclusion}

Using two different methods, our timing observations of PSR~J2019+2425
constrain the orbital inclination to be $i\lesssim 72^\circ$.  Combined with a
model for evolution of the system, this limits the neutron star mass to
$m_1<1.51$\,M$_\odot$, with the median likelihood $m_1=1.33$\,M$_\odot$
for a population~II secondary.  For a population~I secondary the mass
values would be lower.

The use of $\dot{x}$ to place an upper limit on $m_1$ based on a
measurement of $\dot{x}$ has been used for one other pulsar.  PSR~J0437$-$4715
has $\dot{x}=8\times 10^{-12}$, $x=3.367$\,lt-s, $\mu=141$\,mas/yr, which
gives a limit $i<43^\circ$ \cite{sbm+97}.  When combined with the $P_b-m_2$
relation of Rappaport {\it et al.}, this gives $m_1<1.51$\,M$_\odot$
\cite{tc99}, coincidentally the same value as we find for PSR~J2019+2425.

Our finding adds to the growing body of evidence that neutron stars in
neutron-star--white dwarf orbits are not much more massive than those in
neutron star--neutron star binaries.  (See Thorsett
\& Chakrabarty 1999 for a comprehensive review of pulsar mass
measurements\nocite{tc99}).  It is somewhat surprising that the
neutron stars in these systems are not more massive, as the
secondaries in these systems must lose several tenths of a solar mass
as they evolve towards white dwarfs.  Much of this mass could, in
principle, be accreted onto the neutron stars during the low-mass
X-ray binary (LMXB) phase.
To keep the neutron star mass low, most of the
transferred mass must leave the system, perhaps being ejected from the
vicinity of the neutron star via the ``propeller effect'' \cite{is75}.
Taam, King \& Ritter (2000)\nocite{tkr00} suggest 
that wide ($P_{\rm b}$ $\gtrsim$ 2 days) LMXBs are likely to be
transient, with accretion occurring in short, super-Eddington
outbursts that expel most of the transferred mass in a wind.  Low
neutron star masses such as that of PSR~J2019+2425 lend strong
support to this scenario.

\acknowledgements

The Arecibo Observatory is part of the National Astronomy and Ionosphere
Center, which is operated by Cornell University under a cooperative agreement
with the National Science Foundation.  Pulsar research at Princeton University
is supported by National Science Foundation grant AST96-18357.  IHS received
support from an NSERC postdoctoral fellowship.  We thank
F.~Camilo, T.~Tauris, and S.~Thorsett for useful comments on the manuscript;
A.~V\'azquez, F.~Camilo, A.~Lommen, and K.~Xilouris, among others, for help
with data collection; and J.~Taylor for support throughout this project.


\end{document}